\documentclass[10pt,a4paper,english,aps]{revtex4}
\usepackage[T1]{fontenc}
\usepackage[latin1]{inputenc}
\usepackage{babel}
\usepackage{graphics}

\makeatletter

\providecommand{\LyX}{L\kern-.1667em\lower.25em\hbox{Y}\kern-.125emX\@}
\let\SF@@footnote\footnote
\def\footnote{\ifx\protect\@typeset@protect
    \expandafter\SF@@footnote
  \else
    \expandafter\SF@gobble@opt
  \fi
}
\expandafter\def\csname SF@gobble@opt \endcsname{\@ifnextchar[
  \SF@gobble@twobracket
  \@gobble
}
\edef\SF@gobble@opt{\noexpand\protect
  \expandafter\noexpand\csname SF@gobble@opt \endcsname}
\def\SF@gobble@twobracket[#1]#2{}

\usepackage{graphics}

\newcommand{\beq}{\begin{equation}}
\newcommand{\eeq}{\end{equation}}
\newcommand{\bea}{\begin{eqnarray}}
\newcommand{\eea}{\end{eqnarray}}

\def\laq{~\raise 0.4ex\hbox{$<$}\kern -0.8em\lower 0.62
ex\hbox{$\sim$}~}
\def\gaq{~\raise 0.4ex\hbox{$>$}\kern -0.7em\lower 0.62
ex\hbox{$\sim$}~}

\def \pa {\partial}
\def \ra {\rightarrow}
\def \la {\lambda}

\def \Da {\Delta}
\def \b {\beta}
\def \a {\alpha}

\def \ga {\gamma}
\def \sg {\sigma}
\def \da {\delta}

\def \r {\rho}

\def \Om {\Omega}

\def \wt {\widetilde}

\def \hp {\widehat \phi}
\def \hpd {{\dot{\hp}}}

\makeatother
\begin{document}
{\raggedleft BA-TH/02-445\\
 astro-ph/0208032 \par}

\title{Early acceleration and adiabatic matter perturbations \\
 in a class of dilatonic dark-energy models }

\author{L. Amendola\( ^{1} \), M. Gasperini\( ^{2,3} \), D. Tocchini-Valentini\( ^{1} \)
and C. Ungarelli\( ^{4} \) }

\affiliation{\textit{\small \( ^{1} \) Osservatorio Astronomico di Roma, Via
Frascati 33, 00040 Monte Porzio Catone, Italy }\\
\textit{\small \( ^{2} \) Dipartimento di Fisica, Universit\`{a}
di Bari, Via G. Amendola 173, 70126 Bari, Italy }\\
 \textit{\small \( ^{3} \) Istituto Nazionale di Fisica Nucleare,
Sezione di Bari, Bari, Italy }\\
 \textit{\small \( ^{4} \) Institute of Cosmology and Gravitation,
University of Portsmouth, Portsmouth P01 2EG, England }}

\begin{abstract}
We estimate the growth of matter perturbations in a class of recently
proposed dark-energy models based on the (loop-corrected) gravi-dilaton
string effective action, and characterized by a global attractor epoch
in which dark-matter and dark-energy density scale with the same effective
equation of state. Unlike most dark-energy models, we find that the
accelerated phase might start even at redshifts as high as \( z\approx 5 \)
(thus relaxing the coincidence problem), while still producing at
present an acceptable level of matter fluctuations. We also show that
such an early acceleration is not in conflict with the recently discovered
supernova SN1997ff at \( z\approx 1.7 \). The comparison of the predicted
value of \( \sigma _{8} \) with the observational data provides interesting
constraints on the fundamental parameters of the given model of dilaton-dark
matter interactions. 
\end{abstract}
\maketitle

\section{{\small Introduction}}

{\small \label{Sec1}}{\small \par}

{\small Independent cosmological observations have recently pointed
out the existence of a significant fraction of critical density in
the form of unclustered matter, possibly characterized by a negative
pressure \cite{rie,perl,lee,net}. Such a component of the cosmic
fluid, conventionally denoted dark energy, is believed to drive the
accelerated evolution of our present Universe, as suggested by the
study of the Supernovae Ia (SNIa) Hubble diagram. While the simplest
explanation of the dark-energy component is probably a cosmological
constant, it might be neither the most motivated, nor the best fit
to the data. }{\small \par}

{\small The existence of a (presently dominating) dark-energy component
raises at least three important questions:} \emph{\small a}{\small )
what is the physical origin of such a new component?} \emph{\small b}{\small )
why its energy density today is just comparable to the matter energy
density?} \emph{\small c}{\small ) did the acceleration start only
very recently in the cosmological history? }{\small \par}

{\small The first question refers to the fundamental physical mechanism
able to generate a cosmic and universal dark-energy distribution.
So far, most models of dark energy have adopted a purely phenomenological
approach, even if a few proposal concerning the possible role of the
dark-energy field in the context of fundamental physics have already
appeared \cite{fri,wet90,albre,MG01,GPV01,pie}. }{\small \par}

{\small The other questions concern two distinct (although possibly
related) aspects of the so-called {}``coincidence problem{}'' \cite{Ste}.
The first aspect refers to the density of the dark-energy and of the
dark-matter component. Since in most models the two components have
different equations of state, they scale with time in a different
way, and their densities should be widely different for essentially
all the cosmological history: by contrast, current data are strongly
pointing at a comparable proportion of the two components just at
the present epoch. }{\small \par}

{\small The second aspect of the coincidence problem may arise if
the epoch of accelerated evolution started only very recently, say
at \( z\approx 1 \). There are reasons to suspect that this may be
the case because, in most models, the growth of structures is forbidden
after the beginning of the accelerated expansion, so that the acceleration
cannot be extended too far into the past. It has also been recently
claimed that the supernova SN1997ff \cite{super,super1} at \( z\approx 1.7 \)
provides an additional indication that the acceleration is a relatively
recent phenomenon. }{\small \par}

{\small A promising approach for a possible simultaneous answer to
the above questions has been recently provided by a dilatonic interpretation
of the dark energy based on the infinite bare-coupling limit of the
superstring effective action \cite{GPV01}, whose cosmological solutions
are characterized by a late-time global attractor where dark-matter
and (dilatonic) dark-energy densities have an identical scaling with
time. A very similar cosmological scenario has been studied also in
\cite{amtoc,bias}. The respective amount of dark-matter and dark-energy
density is eventually determined by the fundamental constants of the
model, and it is expected to be of the same order, so that their ratio
keeps frozen around unity for the whole duration of such an asymptotic
regime. In such a context, the today approximate equality of dark-matter
and dark-energy density would be no longer a coincidence of the present
epoch, but a consequence of the fact that our Universe has already
entered the asymptotic regime (actually, for a region of parameter
space allowed by observations, it is also possible that the ratio
of dark-energy to dark-matter density is close to unity not only in
the asymptotic regime, but already after the equivalence epoch). Finally,
the cosmic field responsible for the observed large-scale acceleration
would be no longer introduced} \emph{\small ad hoc}{\small , being
identified with a fundamental ingredient of superstring/M-theory models
of high-energy physics. }{\small \par}

{\small In this paper we will focus our discussion on the possible
constraints imposed by structure formation on the above model of dilatonic
dark energy, by studying the growth of matter perturbations in the
two relevant post-equivalence epochs: a first decelerated epoch, dubbed}
\emph{\small dragging phase}{\small , and a second accelerated epoch,
dubbed} \emph{\small freezing phase}{\small . We will then reconstruct
the behavior of the cosmological gravitational potential, evaluate
the Sachs-Wolfe and integrated Sachs-Wolfe contributions to the spectrum
of CMB anisotropies, and estimate the present level of matter fluctuations
(in particular, the so-called variance \( \sigma _{8} \), smoothed
out over spheres of radius 8 Mpc \( h^{-1} \)). Finally, we will
compare it with observations. }{\small \par}

{\small The results that we find are interesting, and might help answering
the third question posed at the beginning of this section: among the
range of parameters compatible with a phenomenologically acceptable
\( \sigma _{8} \) we find indeed values allowing a long epoch of
acceleration, starting as far in the past as at \( z\approx 5 \).
By contrast, the models of dark energy uncoupled to dark matter, with
frozen (or slowly varying) equation of state, cannot accelerate before
\( z\approx 1 \) . It seems appropriate to anticipate here that the
production of an acceptable level of fluctuations even in the case
of an early start of the accelerated epoch is due to two concurrent
factors: the first is that, during the freezing phase, perturbations
do not stop growing like in other models of accelerating dark energy;
the second is that the horizon at equivalence in our model shifts
at larger scales with respect to a standard \( \Lambda  \)CDM model.
It is also important to stress that such an early acceleration is
by no means in contrast with the recently observed supernova SN1997ff
at \( z\approx 1.7 \). }{\small \par}

{\small For the model of dilatonic dark energy considered in this
paper the coincidence problem can thus be alleviated by the fact that
the ratio of dark-matter to dark-energy density is of order one not
only at present, not only in the course of the future evolution but
(at least in principle) also in the past, long before the present
epoch. It should be clear, however, that this possibility could be
strongly constrained by future observations of supernovae at high
redshift. }{\small \par}

{\small The paper is organized as follows. In Sect. \ref{Sec2} we
present the details of our late-time, dilaton-driven cosmological
scenario, and define the (theoretical and phenomenological) parameters
relevant to our computation. In Sect. \ref{Sec3} we discuss the growth
of matter perturbations and compute the variance \( \sg _{8} \).
In Sect. \ref{Sec4} we impose the observational constraints on our
set of parameters, and determine the maximal possible extension towards
the past of the phase of accelerated evolution. We also find, as a
byproduct of our analysis, interesting experimental constraints on
the fundamental parameters of the string effective action used for
a dilatonic interpretation of the dark-energy field. In Sect. \ref{Sec4bis}
we compare our model with the constraints provided by the farthest
observed supernova at \( z\approx 1.7 \). Our conclusions are finally
summarized in Sect. \ref{Sec5}. }{\small \par}

\section{{\small The model}}

{\small \label{Sec2}}{\small \par}

{\small The model we consider is based on the gravi-dilaton string
effective action \cite{GSW87} which, to lowest order in the higher-derivative
expansion, but including dilaton-dependent loop corrections and a
non-perturbative potential, can be written in the String-frame as
follows \cite{GPV01}\begin{equation}
\label{21}
S=-{1\over 2\la _{s}^{2}}\int d^{4}x\sqrt{-\wt g}\left[ e^{-\psi (\phi )}\widetilde{R}+Z(\phi )\left( \wt \nabla \phi \right) ^{2}+{2\la _{s}^{2}}\wt V(\phi )\right] +S_{m}(\wt g,\phi ,{{\textrm{matter}).}}
\end{equation}
 Here \( \la _{s} \) is the fundamental string-length parameter,
and the tilde refers to the String-frame metric. The functions \( \psi (\phi ),Z(\phi ) \)
are appropriate {}``form factors\char`\"{} encoding the dilatonic
loop corrections and reducing, in the weak coupling limit, to the
well known tree-level expressions \cite{GSW87} \( Z=\exp (-\psi )=\exp (-\phi ) \). }{\small \par}

{\small In this paper we are interested in a {}``late-time\char`\"{},
post-big bang cosmological scenario, in which the dilaton is free
to run to infinity rolling down an exponentially suppressed potential,
and the loop form factors approach a finite limit as \( \phi \ra +\infty  \).
Assuming the validity of an asymptotic Taylor expansion \cite{Ven01},
and following the spirit of {}``induced-gravity\char`\"{} models
in which the gravitational and gauge couplings saturate at small values
because of the large number (\( N\sim 10^{2} \)) of fundamental GUT
gauge bosons entering the loop corrections, we can write, for \( \phi \ra +\infty  \),
\begin{eqnarray}
 &  & e^{-\psi (\phi )}\, =\, c_{1}^{2}+b_{1}e^{-\phi }+{{\mathcal{O}}}(e^{-2\phi }),\; Z(\phi )\, =\, -c_{2}^{2}+b_{2}e^{-\phi }+{{\mathcal{O}}}(e^{-2\phi })\; ,\nonumber \\
 &  & \wt {V}(\phi )\, =\, V_{0}\, e^{-\phi }+{{\mathcal{O}}}(e^{-2\phi }).\label{22} 
\end{eqnarray}
 The dimensionless coefficients \( c_{1}^{2},c_{2}^{2} \) are typically
of order \( 10^{2} \), because of their quantum-loop origin. We may
note, in particular, that \( c_{1}^{2} \) asymptotically controls
the fundamental ratio between the (dimensionally reduced) string and
Planck scales \cite{GPV01}, \( c_{1}^{2}=\la _{s}^{2}/\la _{P}^{2} \),
which is indeed expected to be in the range \cite{Kap85} \( \la _{P}/\la _{s}\simeq 0.3-0.03 \). }{\small \par}

{\small To complete the model we have to specify the matter action
\( S_{m} \) of eq. (\ref{21}), containing the coupling (possibly
renormalized by loop corrections) of the matter fields to the dilaton.
The variation of \( S_{m} \) with respect to \( \phi  \) defines
the (String-frame) dilatonic charge density \( \wt \sg  \), whose
appearance is a peculiar string theory effect \cite{Gas99}, and represents
the crucial difference from conventional (Brans-Dicke) scalar-tensor
models of gravity. }{\small \par}

{\small For the cosmological scenario of this paper we shall assume,
as in \cite{GPV01}, that \( S_{m} \) contains radiation, baryons
and cold dark matter, and that the dilatonic charge of the dark matter
component switches on at sufficiently large couplings, being proportional
(through a time-dependent factor \( q(\phi ) \)) to its energy density
\( \wt \rho {_{c}} \). Also, \( q(\phi ) \) is assumed to approach
a constant (positive) value \( q_{0} \) as \( \phi \ra +\infty  \),
\begin{equation}
\label{23}
q(\phi )=q_{0}+{{\mathcal{O}}}(e^{-q_{0}\phi }).
\end{equation}
 The dilatonic charge of radiation and of ordinary baryonic matter
are instead exponentially suppressed in the strong coupling regime
\cite{GPV01}, and this guarantees the absence of unacceptably large
corrections to macroscopic gravity, since in the model we are considering
the dilaton is asymptotically massless and leads to long-range scalar
interactions (see however \cite{DPV02} for possible testable violations
of the equivalence principle, and other non-standard effects, possibly
observable in such a context). }{\small \par}

{\small By considering an isotropic, spatially flat metric background,
and a perfect fluid model of matter sources, it is now convenient
to write the cosmological equations for the action (\ref{21}) directly
in the Einstein frame, defined by the conformal transformation \( \wt g_{\mu \nu }=c_{1}^{2}g_{\mu \nu }\exp (\psi ) \).
In the cosmic-time gauge (and in units \( 2\la _{P}^{2}\equiv 16\pi G=1 \))
the equations are \cite{GPV01}\begin{eqnarray}
 &  & 6H^{2}=\rho {+}\rho {_{\phi }},\quad 4\dot{H}+6H^{2}=-p-p_{\phi },\label{24} \\
 &  & k^{2}(\phi )\left( \ddot{\phi }+3H\dot{\phi }\right) +k(\phi )\, k'(\phi )\, \dot{\phi }^{2}+{V}'(\phi )+\frac{1}{2}\left[ {\psi' (\phi )}(\rho _{b}+\rho {_{c}})+q(\phi )\rho {_{c}}\right] =0,\label{25} 
\end{eqnarray}
 where a prime denotes differentiation with respect to \( \phi  \),
and \begin{eqnarray}
k^{2}(\phi )=3\psi ^{\prime 2}-2e^{2\psi }Z, & \quad V=c_{1}^{4}e^{2\psi }\wt V,\quad  & q(\phi )={\sg _{c}/\r {_{c}}},\label{26} \\
\r {_{\phi }}={1\over2 }k^{2}(\phi )\dot{\phi }^{2}+V(\phi ), &  & p_{\phi }={1\over2 }k^{2}(\phi )\dot{\phi }^{2}-V(\phi ),\label{27} \\
\r {=}\r {_{r}}+\r {_{b}}+\r {_{c}}, &  & p=\r {_{r}}/3.\label{28} 
\end{eqnarray}
 We have explicitly separated the radiation, baryon and cold dark
matter components (\( \rho {_{r}},\rho {_{b}},\rho {_{c}} \)), and
introduced the (Einstein-frame) dilatonic charge per unit of gravitational
mass, \( q(\phi ) \), which is non-vanishing (at large enough \( \phi  \))
only for the dark-matter component. The combination of the above equations
leads to the separate energy conservation equations: \begin{eqnarray}
 &  & \dot{\rho }{_{r}}+4H\rho {_{r}}=0,\label{29} \\
 &  & \dot{\rho }{_{b}}+3H\rho {_{b}}-{1\over 2}\dot{\phi }{\psi' }\rho {_{b}}=0,\label{210} \\
 &  & \dot{\rho }{_{c}}+3H\rho {_{c}}-{1\over 2}\dot{\phi }\left( {\psi' }+q\right) \rho {_{c}}=0,\label{211} \\
 &  & \dot{\rho }{_{\phi }}+3H(\rho {_{\phi }}+p_{\phi })+{1\over 2}\dot{\phi }\left[ {\psi' }(\rho _{b}+\rho {_{c}})+q\rho {_{c}}\right] =0.\label{212} 
\end{eqnarray}
}{\small \par}

{\small With the above assumptions on \( \psi ,Z \) and \( q \),
and for appropriate values of the parameters of the loop functions
(in particular, for a sufficiently small value of \( V_{0} \)), it
has been shown in \cite{GPV01} that the phase of standard, matter-dominated
evolution is modified by the non-minimal, direct coupling of dark
matter and dilatonic dark energy. After the equivalence epoch, in
particular, the Universe may enter a phase of {}``dragging\char`\"{},
followed by an accelerated phase of asymptotic {}``freezing\char`\"{}.
For the reader's convenience we recall here the main properties of
such two phases, referring to \cite{GPV01,amtoc} for a more detailed
discussion (see also \cite{ame3} for a general study of the dynamical
system). }{\small \par}

{\small (1)} \underbar{\small {}``Dragging\char`\"{} phase}{\small .
The potential \( V(\phi ) \) is negligible, the evolution is decelerated,
\( \rho {_{\phi }} \) is still subdominant (as well as \( \rho {_{b}} \)),
but \( \rho {_{\phi }} \) evolves in time like the dominant component
\( \rho {_{c}} \), so that the dilaton dark energy is {}``dragged\char`\"{}
along by the dark matter density. }{\small \par}

{\small In this phase \( k(\phi ) \) can be approximated by a constant,
\( k(\phi )=k_{1} \), and it is thus convenient to define the rescaled
field \( \hp =k_{1}\phi  \) which has canonical kinetic term in the
action, and which satisfies with \( \rho {_{c}} \) the system of
coupled equations \begin{eqnarray}
 &  & {\ddot{\hp }}+3H{\dot{\hp }}+{1\over 2}\left( \psi' +q\over k\right) \rho {_{c}}=0,\label{213} \\
 &  & \dot{\rho }{_{c}}+3H\rho {_{c}}-{1\over 2}\left( \psi' +q\over k\right) \rho {_{c}}{\dot{\hp }}=0,\label{214} 
\end{eqnarray}
 We can then define the canonical effective coupling of the dilaton
to dark matter by the function \( \beta (\phi ) \), defined by: \begin{equation}
\label{215}
\frac{1}{\sqrt{3}}\beta (\phi )={1\over 2}{\psi' (\phi )+q(\phi )\over k(\phi )}
\end{equation}
 which, in the dragging phase, is also approximated by a constant,
\( \beta (\phi )=\beta _{1}\ll1  \) (the conventional factor \( 1/\sqrt{3} \)
has been introduced here to adapt the notations of this paper to previous
studies of the dark-matter-scalar system \cite{ame3}). }{\small \par}

{\small Using eqs. (\ref{24}), (\ref{213}) we find, in the dragging
phase \cite{GPV01}, \begin{equation}
\label{216}
\hpd =-4H\beta _{1}/\sqrt{3},
\end{equation}
 so that, from eq. (\ref{214}), \begin{equation}
\label{217}
\rho {_{c}}\sim H^{2}\sim \rho {_{\phi }}\sim a^{-(3+4\beta _{1}^{2}/3)}.
\end{equation}
 Because of the dragging the time evolution of the dark matter density
deviates from the typical behaviour of dust sources, in such a way
that \( \r {_{c}} \) decays slightly faster than energy density of
baryons, \( \r {_{b}}\sim a^{-3} \). It is however unlikely that
this effect may lead the Universe to a baryon-dominated phase, because
this trend is soon inverted in the subsequent, freezing phase. }{\small \par}

{\small (2)} \underbar{\small {}``Freezing\char`\"{} phase}{\small .
The asymptotic dilaton potential \( V(\phi )=V_{0}\exp (-\phi ) \)
comes into play, the evolution (for large enough values of \( q \))
is accelerated, the dark matter, potential and (dilatonic) kinetic
energy densities evolve in the same way, so that the ratio \( \rho {_{\phi }}/\rho {_{c}} \)
is frozen to an arbitrary constant value. The critical fraction of
dark matter, potential and kinetic energy densities are also separately
constant throughout this phase \cite{GPV01,ame3}. }{\small \par}

{\small In such a phase, asymptotically approached when \( \phi \ra +\infty  \),
one has \( q(\phi )=q_{0} \), \( \psi' =0 \), and the parameters
\( k(\phi ),\beta (\phi ) \) can be again approximated by constant
values \( k_{2},\beta _{2} \) (in general different from the previous
ones), related by \( k_{2}\beta _{2}=\sqrt{3}q_{0}/2 \). The coupled
equations for the canonically rescaled field \( \hp =k_{2}\phi  \)
are modified by the presence of the potential \begin{eqnarray}
 &  & {\ddot{\hp }}+3H{\dot{\hp }}+{\pa V\over \pa \hp }+\frac{\beta _{2}}{\sqrt{3}}\rho {_{c}}=0,\label{218} \\
 &  & \dot{\rho }{_{c}}+3H\rho {_{c}}-\frac{\beta _{2}}{\sqrt{3}}\rho {_{c}}{\dot{\hp }}=0,\label{219} 
\end{eqnarray}
 and, together with eq. (\ref{24}), are solved by the following configuration
\cite{GPV01}: \begin{equation}
\label{220}
V=V_{0}e^{-\hp /k_{2}}\sim (\hpd )^{2}\sim \rho {_{\phi }}\sim \rho {_{c}}\sim H^{2}\sim a^{-6/(2+q_{0})}.
\end{equation}
 In units of critical energy density we find, in particular, \begin{eqnarray}
\Om _{V}={V\over 6H^{2}}=\Om _{k}+{q_{0}\over 2+q_{0}}, &  & \Om _{k}={(\hpd )^{2}\over 12H^{2}}={3k_{2}^{2}\over (2+q_{0})^{2}},\nonumber \\
\Om _{\phi }=\Om _{V}+\Om _{k}, &  & \Om _{c}=1-\Om _{\phi }.\label{221} 
\end{eqnarray}
 Note that, from eq. (\ref{220}), \begin{equation}
\label{222}
{\ddot{a}\over aH^{2}}=1+{\dot{H}\over H^{2}}={q_{0}-1\over q_{0}+2},
\end{equation}
 so that the freezing phase is accelerated for \( q_{0}>1 \). In
that case, according to eq. (\ref{220}), the dark matter density
tends to be strongly enhanced (as time goes on) with respect to the
baryon density, which on the contrary is uncoupled to the dilaton
and thus evolves in the standard way, \( \rho {_{b}}\sim a^{-3} \).
As already pointed out in \cite{GPV01,AT01a}, it is tempting to speculate,
in such a context, that the present smallness of the ratio \( \rho {_{b}}/\rho {_{c}} \)
could then emerge as an artifact of a long enough freezing phase,
started before the present epoch. }{\small \par}

{\small In this paper, in order to discuss the possible bounds imposed
by present observations on the above scenario, we will consider a
simplified model of late-time (i.e., after-equivalence) cosmology,
consisting of two phases. More precisely, we will drastically approximate
the background evolution by assuming that the Universe performs a
sudden transition from the radiation-dominated to the dragging phase
at the equivalence epoch \( a=a_{{\textrm{eq}}} \), and from the
dragging to the freezing phase at the transition epoch \( a=a_{f} \).
We will discuss in this context the phenomenological constraints on
the parameters of the string effective action, and in particular their
possible consistency with an early beginning of the freezing epoch,
\( z_{f}=(a_{0}/a_{f})-1\gg1  \), which (as already mentioned) may
be relevant for a truly satisfactory solution of the coincidence problem. }{\small \par}

{\small To make contact with previous results we will use here the
following explicit model of dilaton potential, charge and loop corrections
(already adopted in \cite{GPV01}): \begin{eqnarray}
&&e^{-\psi (\phi )}\, =\, e^{-\phi }+c_{1}^{2}, \quad Z(\phi )\, =\, e^{-\phi }-c_{2}^{2},\quad q(\phi )\, =q_{0}\, {e^{q_{0}\phi }\over c^{2}+e^{q_{0}\phi }},\label{223} \\
&&V(\phi )\, \, =c_{1}^{4}\, \, m_{V}^{\, 2}e^{2\psi }\left[ \exp \, (-e^{-\phi }/\a _{1})-\exp \, (-e^{-\phi }/\a _{2})\right] , \quad 0<\a _{2}<\a _{1}.\label{224} 
\end{eqnarray}
 Here \( c_{1}^{2},c_{2}^{2},c^{2} \) are numbers of order \( 10^{2} \),
the asymptotic charge satisfies \( q_{0}>1 \) to guarantee a final
accelerated regime, and the dilaton potential reduces asymptotically
to the exponential form of eq. (\ref{22}), with \( V_{0}=m_{V}^{2}(\alpha _{1}-\alpha _{2})/\alpha _{1}\alpha _{2} \)
(in units \( 2\la _{P}^{2}=1 \)). It is worth noting that the mass
scale \( m_{V} \), controlling asymptotically the amplitude of the
potential (and thus the beginning of the freezing phase), is possibly
expected to be of non-perturbative origin, and thus related to the
fundamental string scale in a typical instantonic way, \begin{equation}
\label{225}
m_{V}=\exp \left[ -{2\over \beta ^{*}{\alpha }_{{\textrm{GUT}}}}\right] M_{s},
\end{equation}
 where \( \alpha _{{\textrm{GUT}}}\simeq 1/25 \) is the asymptotic
value of the GUT gauge coupling, and \( \beta ^{*}{} \) is some model-dependent
loop coefficient. As noted in \cite{GPV01}, a value of \( \beta ^{*} \)
slightly smaller than the usual coefficient of the QCD beta-function
is already enough to move \( m_{V} \) from the QCD scale down to
the scale relevant for a realistic scenario of dark-energy domination;
a typical reference value is, for instance, \( \beta ^{*}\simeq 0.36 \),
which corresponds to \( m_{V}\sim H_{0}\sim 10^{-61}M_{P} \), and
thus to a freezing phase starting around the present epoch. }{\small \par}

{\small With the above explicit forms of the loop corrections we can
now compute the constant parameters for the dragging (\( k_{1},\beta _{1} \))
and the freezing (\( k_{2},\beta _{2} \)) eras. By setting \( \mu _{2}=\sqrt{3}c_{1}/(\sqrt{2}c_{2}) \)
we find \begin{equation}
\label{226}
k_{1}={\sqrt{3}\over \mu _{2}},\quad \beta _{1}={q_{0}\mu _{2}\over 2c^{2}},\quad k_{2}={\sqrt{3}\over \mu _{2}},\quad \beta _{2}={q_{0}\mu _{2}\over 2}.
\end{equation}
 The constant \( \mu _{2}/\sqrt{3} \) represents the slope of the
dilaton potential \( V(\hp ) \) during freezing. With these definitions,
\( q_{0}=2\beta {_{2}}/\mu _{2},c^{2}=\beta {_{2}}/\beta {_{1}} \). }{\small \par}

{\small For our model of background we can finally express the phenomenological
variables, required for the subsequent computations, in terms of the
above set of parameters. From eqs. (\ref{217}), (\ref{220}) we get
the barotropic parameter \( w=(p_{tot}/\rho _{tot}{)}+1 \) relative
to the effective equation of state of cold dark matter, \begin{equation}
\label{230}
w_{1}=1+{4\over 9}\beta _{1}^{2},\quad w_{2}={2\over 2+q_{0}}={\mu _{2}\over \mu _{2}+\beta {_{2}}},
\end{equation}
 in the dragging and freezing eras, respectively. Another useful parameter
is \( \Omega _{c} \): in the dragging phase, from eq. (\ref{216}),
\begin{equation}
\label{231}
\Omega _{c}=1-\Omega _{\phi }=1-{(\hpd )^{2}\over 12H^{2}}=1-{4\over 9}\beta {_{1}}^{2}.
\end{equation}
 In the freezing phase, from eq. (\ref{221}), \begin{equation}
\label{232}
\Omega _{c}={2\mu _{2}^{2}+2\beta {_{2}}\mu _{2}-9\over 2(\beta {_{2}}+\mu _{2})^{2}}.
\end{equation}
(this is therefore to be identified to the value observed today).}{\small \par}

{\small Note that the ratio of dark-energy to dark-matter density
may be close to unity throughout the cosmic evolution after equivalence,
if the constants \( \beta _{1},\beta _{2},\mu _{2} \) are of the
order of unity. In this sense, in such a model, it is even possible
that a serious coincidence problem never really arises. By contrast,
dark-energy models} \emph{\small without} {\small a stationary phase
in which \( \r {_{m}}\sim \r {_{\phi }} \) (see e.g. \cite{cald,kessence})
have to explain the today value (or order one) of ratios \( \Omega _{\phi }/\Omega _{m} \)
which range from extremely small values in the past to unboundedly
large values in the future \cite{amtoc}. }{\small \par}

{\small We end this section with the computation of three useful red-shift
parameters \( z_{e},z_{f},z_{b} \) corresponding, respectively, to
the radiation-matter equivalence, to the beginning of the freezing
epoch, and to the baryon epoch (associated to a possible baryon-dominated
phase). From the behaviour of \( \rho {_{\phi }} \) and \( \rho {_{b}} \)
in the freezing phase, \begin{equation}
\label{233}
\rho {_{\phi }}=\rho {_{\phi }}(a_{0})\left( a_{0}\over a\right) ^{3w_{2}},\quad \rho {_{b}}=\rho {_{b}}(a_{0})\left( a_{0}\over a\right) ^{3},
\end{equation}
 we can determine the baryon red-shift epoch \( a_{b} \), such that
\( \rho {_{\phi }}=\rho {_{b}} \), as follows \cite{AT01a}: \begin{equation}
\label{234}
1+z_{b}={a_{0}\over a_{b}}=\left( \Omega _{\phi }\over \Omega _{b}\right) _{0}^{\mu _{2}+\beta {_{2}}\over 3\beta {_{2}}}.
\end{equation}
 The present ratio \( (\Omega _{\phi }/\Omega _{b})_{0} \) is a known
observational input. Our model excludes the possibility of a baryon-dominated
phase and thus requires, for consistency, \( z_{b}>z_{f} \) (see
\cite{AT01a} for a dark-energy model with a baryonic epoch). }{\small \par}

{\small The equivalence scale can be obtained by rescaling \( \rho {_{c}} \)
and \( \rho {_{r}} \) from \( a_{e} \) down to \( a_{0} \), i.e.
\begin{equation}
\label{235}
\rho {_{c}}(a_{0})=\rho {_{c}}(a_{e})\left( a_{e}\over a_{f}\right) ^{3w_{1}}\left( a_{f}\over a_{0}\right) ^{3w_{2}},\quad \rho {_{r}}(a_{0})=\rho {_{r}}(a_{e})\left( a_{e}\over a_{f}\right) ^{4}\left( a_{f}\over a_{0}\right) ^{4},
\end{equation}
 from which \begin{equation}
\label{236}
1+z_{e}={a_{0}\over a_{e}}=\left[ \left( \Omega _{r}\over \Omega _{c}\right) _{0}\left( a_{f}\over a_{0}\right) ^{3(w_{2}-w_{1})}\right] ^{1\over 3w_{1}-4},
\end{equation}
 where, again, \( (\Omega _{r}/\Omega _{c})_{0} \) is an observational
input. Similarly, we can determine the freezing epoch by rescaling
the dilaton potential energy, \begin{equation}
\label{237}
\rho {_{V}}(a_{f})=V_{0}e^{-\phi _{f}}=\rho {_{V}}(a_{0})\left( a_{0}\over a_{f}\right) ^{3w_{2}}.
\end{equation}
 Here \( \rho {_{V}}(a_{0})=6H_{0}^{2}\Omega _{V} \), where \( \Omega _{V} \)
is determined by eq. (\ref{221}), and \( \phi _{f} \) is the transition
scale between small and large values of the dilaton charge, namely
\( \phi _{f}=(2/q_{0})\ln c \), from eq. (\ref{223}). We thus obtain
\begin{eqnarray}
1+z_{f}={a_{0}\over a_{f}} & = & \left[ {V_{0}\over 6H_{0}^{2}\Omega _{V}}\left( \beta {_{1}}\over \beta {_{2}}\right) ^{\mu _{2}\over 2\beta {_{2}}}\right] ^{\mu _{2}+\beta {_{2}}\over 3\mu _{2}},\\
\Omega _{V} & = & {9+4\beta {_{2}}(\mu _{2}+\beta {_{2}})\over 4(\mu _{2}+\beta {_{2}})^{2}},\label{238} 
\end{eqnarray}
 where \( H_{0} \) is the value of the Hubble parameter provided
by present observations. }{\small \par}

{\small In conclusion, we have a four-parameter model of background,
spanned by two possible (related) sets of independent variables: the
phenomenological variables \( \{\beta {_{1}},\beta {_{2}},\mu _{2},z_{f}\} \)
or, equivalently, the fundamental variables \( \{c_{1}/c_{2},c^{2},q_{0},V_{0}\} \),
referred to the \( \phi \ra +\infty  \) limit of the string effective
action of eq. (\ref{21}). The mapping between the two sets is defined
by eqs. (\ref{226}) and (\ref{238}). Note that two of the four parameters
can be in principle determined by fitting the observational data relative
to the present fraction of cold dark matter, \( \Om _{c} \), and
the dark-energy equation of state, \( w_{2} \): we can determine,
for instance, \( \mu _{2} \) and \( \b {_{2}} \) through eqs. (\ref{226}),
(\ref{232}). In the following sections we will discuss the allowed
regions left by various phenomenological constraints in such a parameter
space. }{\small \par}

\section{{\small Constraints from structure formation}}

{\small \label{Sec3}}{\small \par}

{\small In this Section we will compute the r.m.s. dark-matter density
contrast \( \sigma _{8} \) for the model presented in Sect. \ref{Sec2},
combining it with the CMB angular power spectrum (at low multipoles)
in order to eliminate the dependence on the normalization factor.
Another analytical computation of \( \sigma _{8} \) including dark
energy has recently been presented in \cite{doran e schwindt}, but
only for models in which dark energy and dark matter are uncoupled. }{\small \par}

{\small Before starting the computation, let us recall some preliminary
condition to be imposed (for consistency) on our model of background. }{\small \par}

\subsection{{\small Consistency conditions on the background}}

{\small \label{Sec3a}}{\small \par}

{\small As illustrated in Sect. \ref{Sec2}, the evolution of our
cosmological background is characterized by two stationary (\( \rho _{c}\sim \rho _{\phi } \))
stages: the dragging era (labelled by the subscript \( 1 \)), and
the final accelerated freezing era (labelled by \( 2 \)). This scenario
can be consistently implemented provided the parameters are chosen
in such a way as to satisfy the following conditions. }{\small \par}

{\small First, the dragging era exists, and is a saddle point of our
dynamical system, only if \cite{ame3}: \begin{equation}
\left| \beta _{1}\right| <{\sqrt{3}}/{2}\simeq 0.87.
\end{equation}
 Second, in order for the baryons not to dominate (in the past) over
the dark-matter component, it is necessary to impose (as anticipated):
\begin{equation}
\label{barfreez}
z_{b}>z_{f}.
\end{equation}
 Third, the final freezing phase is accelerated only if: \begin{equation}
\label{accel}
\mu _{2}<2\beta _{2}.
\end{equation}
 The current SNIa observations, however, require more than simply
an acceleration, as shown by a recent analysis \cite{Dalal} of SNIa
data in models which include a freezing epoch. The result is that
only models with an effective equation of state \( w_{2}\approx 0.4 \)
(best fit) or \( w_{2}<0.5 \) (at one sigma) are consistent with
the SNIa Hubble diagram. }{\small \par}

{\small In the following discussion we will reduce, for simplicity,
our set of free parameters by using the observed CDM fraction of critical
density, \( \Om _{c}\simeq 0.3 \), as a fixed observational input.
This can be used to eliminate \( \mu _{2} \) through eq. (\ref{232}),
in such a way that we are left with three parameters only, \( \b {_{1}},\b {_{2}} \)
and \( z_{f} \). In that case, the above limits on \( w_{2}=\mu _{2}/(\mu _{2}+\beta _{2}) \)
define two reference values for \( \beta _{2} \), \begin{eqnarray}
\beta _{2} & = & 4.02,\: {(\textrm{best fit })}\\
\beta _{2} & = & 2.35,\: {(\textrm{one sigma })},\label{min} 
\end{eqnarray}
 which will be used in our subsequent analysis. For future reference,
note that the lower limit at 95\% c.l. is \( \beta _{2}>1.55 \).
We note, finally, that pushing back in time the transition between
the dragging and freezing eras, that is increasing \( z_{f}, \) implies
a decrease of \( z_{e} \). The condition \begin{equation}
z_{f}\laq 100
\end{equation}
 prevents the unwanted crossing of these two quantities. }{\small \par}

{\small All the above constraints will be imposed on our subsequent
computations. }{\small \par}

\subsection{{\small Angular power spectrum at low multipoles}}

{\small In this subsection we will extend the treatment of the Sachs-Wolfe
effect presented in \cite{HuW} (to which we refer for the notation)
in such a way as to include the case of coupled dark energy and dark
matter. We shall assume a conformally flat metric, \( ds^{2}=a^{2}(d\tau ^{2}-dx^{i}dx_{i}) \),
and we shall exploit the fact that, in the absence of anisotropic
stress, the two scalar potentials \( \Psi  \) and \( \Phi  \) defined
in the longitudinal gauge turn out to be equal and to coincide, in
the Newtonian limit, with the usual gravitational potential. The general
expressions for the Sachs-Wolfe (SW) and integrated Sachs-Wolfe (ISW)
parts of the angular power spectrum, for adiabatic scalar perturbations,
can then be written as \cite{HuW}: \begin{eqnarray}
C_{\ell }^{SW} & = & \frac{2}{9\pi }\int dkk^{2}\left| \Psi (k,\tau _{d})\right| ^{2}j_{\ell }^{2}[k(\tau _{0}-\tau _{d})],\\
C_{\ell }^{ISW} & = & 2G^{2}_{2}(\ell )\int dk\left| {\Psi }'(k,\tau _{\ell })\right| ^{2},
\end{eqnarray}
 where \( G^{}_{2}(\ell )={\Gamma [(\ell +1)/2]}/{\Gamma [(\ell +2)/2]} \),
\( j_{\ell } \) are the spherical Bessel functions, the prime denotes
differentiation with respect to conformal time, and \( \tau _{d} \)
is the conformal time at decoupling, while \( \tau _{\ell }=\tau _{0}-(\ell +1/2)/k \)
(notice that the above expression for the ISW coefficient has been
already integrated over \( \tau  \)). }{\small \par}

{\small Let us denote with \( \delta _{k} \) the CDM density contrast
for the wavemode \( k \), and with \( \hp  \) and \( \chi \equiv \delta \hp  \),
respectively, the values of the background scalar field and of its
fluctuation. The label \( 0 \) will denote the present time \( t_{0} \),
and it is to be understood that all the density parameters \( \Omega _{c},\Omega _{b} \),
etc. are always referred to the present time, unless otherwise stated.
Finally, all perturbation variables will be expressed in terms of
their Fourier components. The potential \( \Psi  \) is then determined
by the relativistic Poisson equation (in units \( 16\pi G=1 \)) \begin{equation}
\label{44}
\Psi (a)=-\frac{3}{2k^{2}}H_{0}^{2}\Omega _{c}\delta _{k}(a_{0})s(k,a)-\frac{1}{4k^{2}}{\chi }'{\hp }'-\frac{3a^{}H^{}}{4k^{2}}\chi {\hp }'-\frac{a^{2}}{4k^{2}}{\chi }\frac{dV}{d{\hp }},
\end{equation}
 where we have introduced the (possibly \( k \)-dependent) function
\( s(k,a) \), which accounts for the time-evolution of the dark matter
and of its density contrast, according to the equations \begin{eqnarray}
\rho _{c}(a) & = & \rho _{c}(a_{0})f(a),\\
\delta _{k}(a) & = & \delta _{k}(a_{0})D(k,a),\\
s(k,a) & = & D(k,a)f(a)\left( \frac{a}{a_{0}}\right) ^{2}.
\end{eqnarray}
}{\small \par}

{\small The quantities appearing in the r.h.s. of eq. (\ref{44})
are evaluated in the synchronous gauge, while the potential \( \Psi  \)
is calculated in the longitudinal gauge \cite{HuW}, since this helps
extending the validity of the Poisson equation to all scales. Note
that we have dropped from the Poisson equation the contribution of
the dark-matter velocity fluctuations, since it can be shown that
they are negligible \cite{bias}. The baryon contribution has been
neglected as well. Defining the matter power spectrum \( P(k)=|\delta _{k}(a_{0})|^{2} \),
it follows that \begin{eqnarray}
C_{\ell }^{SW} & = & \frac{1}{2\pi }H_{0}^{4}\Omega _{c}^{2}(1+\Sigma )\int \frac{dk}{k^{2}}P(k)|s(k,a_{d})|^{2}j_{\ell }^{2}[k(\tau _{0}-\tau _{d})],\label{48} \\
C_{\ell }^{ISW} & = & \frac{9}{2}H_{0}^{4}\Omega _{c}^{2}G^{2}_{2}(\ell )(1+\Pi )\int \frac{dk}{k^{4}}P(k)\left| s'[k,a(\tau _{\ell })]\right| ^{2}.
\end{eqnarray}
 The functions \( \Sigma  \) and \( \Pi  \) represent corrections
due to the scalar field contribution. In the next few paragraphs we
will neglect these effects, concentrating the attention only on the
dark-matter part of the power spectrum, while the scalar contribution
will be reconsidered at the end of our calculation. The convenience
of this procedure will become apparent later on. }{\small \par}

{\small Focusing our attention on the background evolution after the
epoch of matter-radiation equivalence, we can now specify the evolution
of the CMB energy density by setting \begin{eqnarray}
f(a) & = & (a/a_{0})^{-3w_{2}},\qquad a>a_{f}\\
 & = & (a_{f}/a_{0})^{-3w_{2}}(a/a_{f})^{-3w_{1}},\qquad a<a_{f}.
\end{eqnarray}
 where \( w_{1,2} \) were defined in eqs. (\ref{230}). For what
concerns the growth of matter perturbations, since we are interested
in the low-multipole branch of the spectrum, it is reasonable to consider
only scales that reenter the horizon after equivalence (and before
freezing, as in the accelerated stage no reenter is possible). The
growth of \( \da _{k} \) has been derived in \cite{bias} as a function
of the parameters \( \b {} \) and \( \mu  \) of the dragging and
freezing epochs. It turns out that the evolution is the same for all
modes: }{\small \par}
{\small 
\begin{eqnarray*}
D(k,a) & = & (a/a_{0})^{m_{2}},\quad a>a_{f}\\
 & = & (a/a_{f})^{m_{1}}(a_{f}/a_{0})^{m_{2}},\quad a<a_{f},
\end{eqnarray*}
 where \begin{equation}
m_{1}=1+\frac{4}{3}\beta ^{2}_{1},\quad m_{2}={\Delta -10\beta _{2}-\mu _{2}\over 4\left( \beta _{2}+\mu _{2}\right) },
\end{equation}
 and \begin{equation}
\Delta ^{2}=-108+44\beta _{2}\mu _{2}+32\beta ^{3}_{2}\mu _{2}+25\mu ^{2}_{2}+\beta ^{2}_{2}\left( 32\mu ^{2}_{2}-44\right) 
\end{equation}
 (in the dragging phase the same exponent \( m_{1} \) describes the
growth of perturbations both inside and outside the horizon). Considering
scales with \( k<k_{d} \), where the subscript \( d \) stands for
decoupling, the relevant function for the computation of the ordinary
SW effect can thus be parametrized by a \( k \)-independent function
as follows: \begin{equation}
s(a_{f})=(a_{d}/a_{f})^{m_{1}-3w_{1}}(a_{f}/a_{0})^{m_{2}-3w_{2}}(a_{d}/a_{0})^{2}=(a_{f}/a_{0})^{\alpha _{2}}.
\end{equation}
where \( \alpha _{2}\equiv m_{2}-3w_{2}+2 \) (note that \( \alpha _{1}\equiv m_{1}-3w_{1}+2=0 \)).}{\small \par}

{\small Some comments are now in order, concerning the evolution of
perturbations described by the above equations. First we notice that
in the dragging phase \( m_{1}>1 \), i.e. that the perturbations
growth is faster than in a standard CDM model. This is due to an extra
pull on dark matter arising from the dark-energy coupling, that act
as an additional (scalar) gravity force \cite{bias}. Secondly, the
growth in the accelerated freezing regime does not vanish asymptotically,
like in other dark-energy models. This is again an effect of the dark
energy-dark matter coupling and of the fact that the dark-matter density
is not driven to zero by the acceleration. Finally, since the baryons
are decoupled from the dilaton, they evolve differently, and a bias
between the baryons and the dark-matter distribution is expected.
The constraints from this effect have been discussed in ref. \cite{bias}. }{\small \par}

{\small It is important to stress that the matter power spectrum,
calculated today for scales \( k<k_{e} \), where \( k_{e} \) is
the scale that reenter the horizon at equivalence, \begin{equation}
\label{eq}
k_{e}=a_{e}H_{e}=a_{e}H_{0}\left[ (a_{f}/a_{0})^{-3w_{2}}(a_{e}/a_{f})^{-3w_{1}}\right] ^{1/2},
\end{equation}
 does not change respect to the primordial shape \begin{equation}
P(k)=Ak^{n},
\end{equation}
 where \( A \) is the usual normalization factor (see e.g. ref. \cite{KT}).
This is due to a peculiarity of the dragging phase, i.e. to the fact
that the perturbation growth is identical inside and outside the horizon,
just like in the case of standard CDM models (although, as already
mentioned, the growth rate \( m_{1} \) is larger than unity). The
dragging phase, in addition, leads to a time-independent gravitational
potential (again, like in standard CDM models), and thus it does not
contribute to the ISW effect, which is entirely produced in the subsequent
freezing phase. }{\small \par}

{\small From the integration of eq. (\ref{48}), with \( \Sigma =0 \)
we then easily obtain \begin{equation}
C_{\ell }^{SW}=\frac{\gamma ^{1-n}}{16}AH_{0}^{n+3}\Omega _{c}^{2}s^{2}(a_{f})G(\ell \, ,n),
\end{equation}
 where \begin{equation}
G(\ell \, ,n)=\frac{\Gamma (3-n)\Gamma ((2\ell +n-1)/2)}{\Gamma ((4-n)/2)^{2}\Gamma ((2\ell +5-n)/2)},
\end{equation}
 and where the factor \( \gamma  \) appears by eliminating \( \tau _{0} \)
(from the result of the SW integral) in terms of the present Hubble
scale \( H_{0} \), according to the definition \begin{equation}
\tau _{0}=\int \frac{da}{a^{2}H}=\frac{2}{a_{0}H_{0}}\gamma .
\end{equation}
 This integral can be finally estimated by considering the separate
contributions from the two phases of our model, and we obtain: \begin{equation}
\gamma =\frac{1}{2\lambda _{2}}\left[ 1+\frac{3\Delta w\left( a_{f}/a_{0}\right) ^{\lambda _{2}}}{2\lambda _{1}}\right] ,\; \lambda _{1}=\frac{3}{2}w_{1}-1,\; \lambda _{2}=\frac{3}{2}w_{2}-1,\; \Da w=w_{2}-w_{1}
\end{equation}
 (the usual result for the standard cosmological model is instead
\( \ga =1 \)). }{\small \par}

{\small It is appropriate to reconsider at this point the scalar-field
corrections to the SW effect, represented by \( \Sigma  \). The scalar
field fluctuations which are outside the horizon during the dragging
phase grow proportionally to the CDM density contrast (see \cite{ame3,AT01a}).
Since \( \r {_{c}}\sim \r {_{\phi }} \), it is found that \( \Sigma  \)
is a constant, and depends only on \( \b {_{1}} \) as follows: \begin{equation}
\label{sigma}
\Sigma =\frac{64\sqrt{\frac{\pi }{3}}\beta ^{2}_{1}(3+\beta ^{2}_{1})(15+4\beta ^{2}_{1})}{405+252\beta ^{2}_{1}-336\beta ^{4}_{1}+64\beta ^{6}_{1}}
\end{equation}
 (the contribution coming from the scalar-field potential has been
neglected, since in the dragging phase the dilaton kinetic energy
is dominant with respect to \( V \)). }{\small \par}

{\small For what concerns the ISW effect, the scalar field contribution
on sub-horizon scales is negligible \cite{AT01a}, and we can consistently
set \( \Pi =0. \) If we define the variable \( y=k\tau _{0} \),
and we use the result that, in the accelerated epoch, \begin{equation}
s(a)=(a/a_{0})^{\alpha _{2}},
\end{equation}
 then we are lead to \begin{equation}
C_{\ell }^{ISW}=\frac{9}{2}A\Omega _{c}^{2}G^{2}_{2}(\ell )H_{0}^{n+3}\alpha _{2}^{2}\int ^{\infty }_{y_{min}}dyy^{n-4}a(\ell ,\, y)^{2\alpha _{2}-2\lambda _{2}}.
\end{equation}
 The lower limit of integration, \begin{equation}
y_{min}=\frac{\left( \ell +\frac{1}{2}\right) }{1-r},\quad r=\frac{\tau _{f}}{\tau _{0}}=\frac{(a_{f}/a_{0})^{\lambda _{2}}}{2\gamma \lambda _{1}},
\end{equation}
 has been obtained by imposing \( \tau _{\ell }>\tau _{f} \), since
only in the second (freezing) stage there is a significant ISW contribution.
The time-dependence of the freezing scale-factor, on the other hand,
can be parametrized by \begin{equation}
a=a_{0}\left[ \lambda _{2}\left( H_{0}\tau +B\right) \right] ^{1/\lambda _{2}},
\end{equation}
 where \(  \)\begin{equation}
B=\left( \frac{a_{f}}{a_{0}}\right) ^{\lambda _{2}}\left( \frac{1}{\lambda _{2}}-\frac{1}{\lambda _{1}}\right) .
\end{equation}
}{\small \par}

{\small By using all the above results the ISW integral can be finally
performed and, in the case \( n=1 \), the result is \begin{equation}
C_{\ell ,ISW}=\frac{9}{4}A\Omega _{c}^{2}G^{2}_{2}(\ell )H_{0}^{4}\alpha _{2}^{}\frac{\left[ (B+2\gamma )\lambda _{2}\right] ^{2\frac{\alpha _{2}}{\lambda _{2}}}+\left[ (B+2\gamma r)\lambda _{2}\right] ^{2\frac{\alpha _{2}}{\lambda _{2}}-1}\cdot \left[ 4\gamma \alpha _{2}(r-1)-(B+2\gamma r)\lambda _{2}\right] }{\gamma ^{2}(2\ell +1)^{2}(2\alpha _{2}-\lambda _{2})}.
\end{equation}
 For a generic, primordial spectral index with \( n\not =1 \) a much
more complicated (but still analytic) expression may be obtained. }{\small \par}

{\small In conclusion, the dimensionless, angular power-spectrum at
low multipoles \( \left( \ell \leq 10\right)  \) can be approximated
by \begin{equation}
C_{\ell }=C_{\ell ,SW}+C_{\ell ,ISW}=AF,
\end{equation}
 where, for \( n=1 \), \begin{eqnarray}
 &  & F\left( \ell ;a_{f},\, \beta _{1},\, \beta _{2},\, \mu _{2}\right) =H_{0}^{4}\Omega _{c}^{2}\Bigg [\frac{1}{4\pi \ell (\ell +1)}\left( \frac{a_{f}}{a_{0}}\right) ^{2\alpha _{2}}\left( 1+\Sigma \right) \nonumber \\
 &  & +\frac{9}{4}^{}G^{2}_{2}(\ell )^{}\alpha _{2}^{}\frac{\left[ (B+2\gamma )\lambda _{2}\right] ^{2\frac{\alpha _{2}}{\lambda _{2}}}+\left[ (B+2\gamma r)\lambda _{2}\right] ^{2\frac{\alpha _{2}}{\lambda _{2}}-1}\cdot \left[ 4\gamma \alpha _{2}(r-1)-(B+2\gamma r)\lambda _{2}\right] }{\gamma ^{2}(2\ell +1)^{2}(2\alpha _{2}-\lambda _{2})}\Bigg ].
\end{eqnarray}
 The variable \( C_{\ell }^{*} \), representing the experimentally
observed angular power spectrum, measured in units of \( (\mu K)^{2} \),
can thus be finally written in the form \begin{equation}
\label{acl}
C_{\ell }^{*}={T_{0}^{2}\ell (\ell +1)AF\over 2\pi },
\end{equation}
 where \( T_{0}=2.726\times 10^{6}\mu  \)K. }{\small \par}

\subsection{{\small Calculation of \protect\protect\( \sigma _{8}\protect \protect \)}}

{\small The (dimensionless) variance of the CMB density fluctuations,
in spheres of radius \( R_{8}=8h^{-1} \) Mpc, where \( h=H_{0}/(100{\textrm{kmsec}}^{-1}{\textrm{Mpc}}^{-1}) \),
is defined by \cite{KT}\begin{equation}
\label{sg8}
\sigma _{8}^{2}=\frac{1}{2\pi ^{2}}\int P(k)W_{8}^{2}(k)k^{2}dk=\frac{A}{2\pi ^{2}}R_{8}^{-(3+n)}I_{1},
\end{equation}
 where \( W_{8}(k) \) is the spherical top-hat window function of
radius \( R_{8} \), and \begin{equation}
\label{inti1}
I_{1}=\int x^{2+n}T^{2}(x)W^{2}(x)dx.
\end{equation}
 Note that we are using the full power spectrum corrected by the transfer
function \( T(k) \), i.e. \( P(k)=Ak^{n}T^{2}(k) \), since in the
definition of \( \sigma _{8} \) it is necessary to include also scales
with \( k>k_{e} \). }{\small \par}

{\small It is important to stress, at this point, that the above transfer
function is identical to the one of the usual \( \Lambda  \)CDM model.
During the dragging phase, in fact, the perturbation growth does not
depend on the wavenumber, while during the freezing phase only sub-horizon
perturbations have to be taken into account, so that no distortion
of the power spectrum occurs after the equivalence epoch. Therefore,
the transfer function only expresses the usual correction to the primordial
spectrum due to the different growth of perturbations in the radiation
epoch (those entering the horizon before equivalence are depressed
with respect to those entering later). Since, in our model, the cosmological
evolution before equivalence is standard, we can safely adopt the
transfer function of a \( \Lambda  \)CDM model} \emph{\small for
which the wavenumber \( k_{e} \) crossing the horizon at equivalence
is the same as in our model}{\small . For a \( \Lambda  \)CDM model
with present density \( \Omega _{c(\Lambda cdm)} \) one has, in particular,
\begin{equation}
k_{e(\Lambda cdm)}=a_{0}H_{0}\Omega _{c(\Lambda cdm)}\sqrt{\frac{2}{\Omega _{r}}}.
\end{equation}
 By equating to \( k_{e(\Lambda cdm)} \) the value of \( k_{e} \)
determined in our model (see eq. \ref{eq}), we thus obtain the effective
density parameter \begin{equation}
\Omega _{c(\Lambda cdm)}=\left( \Om _{r}\over 2\right) ^{1\over 2}\left( a_{f}\over a_{0}\right) ^{-{3\Delta w\over 2}}\left( a_{e}\over a_{0}\right) ^{2-3w_{1}\over 2}
\end{equation}
 to be used for the determination of the equivalent transfer function
(we shall of course restrict our analysis to the case \( \Omega _{c(\Lambda cdm)}<1 \)).
For the final numerical integration of eq. (\ref{inti1}) we shall
use the \( \Lambda CDM \) transfer function proposed in \cite{Hu e Su 96}. }{\small \par}

{\small Combining eq. (\ref{acl}) and eq. (\ref{sg8}), in order
to eliminate the normalization factor \( A \), we finally obtain
\begin{equation}
\label{relazione}
\sigma ^{2}_{8}=\frac{C_{\ell }^{*}I_{1}}{\pi T_{0}^{2}R^{4}_{8}\ell (\ell +1)F},
\end{equation}
 where \( I_{1} \) depends on \( \Omega _{c\, (\Lambda cdm)} \).
The observed \( C^{*}_{\ell } \) has been obtained by fitting the
COBE data as in \cite{Bunn}. The comparison of eq. (\ref{relazione})
with the experimental value of \( \sg _{8} \), \begin{equation}
\label{exp}
\sigma _{8}=(0.56\pm 0.1)\Omega ^{-0.47}_{c},
\end{equation}
 taken from data on clusters abundances \cite{vialid}, will eventually
give the sought-for constraint on the parameters of the dilaton model
introduced in Sect. \ref{Sec2}. }{\small \par}

\section{{\small Results}}

{\small \label{Sec4}}{\small \par}

{\small In order to implement the constraints imposed by \( \sg _{8} \)
we shall first eliminate \( \mu _{2} \) from our set of parameters
(as already anticipated), by using eq. (\ref{232}) and fixing the
CDM density at the value \( \Om _{c}=0.3 \), suggested by present
observation (see e. g. \cite{lahav}). Also, we shall assume for the
baryon density the value predicted by standard nucleosynthesis \cite{burles},
\( \Omega _{b}=0.02\, h^{-2} \), and we set \( h=0.65 \). }{\small \par}

{\small The value of \( \b {_{2}} \), at given \( \mu _{2} \), should
be unambiguously determined by \( q_{0} \), and then by the observed
value of the cosmic acceleration through eqs. (\ref{222}) and (\ref{226}).
However, in view of the present experimental uncertainties, we have
accepted here an open range of possibilities and we have illustrated
the constraints at two (rather different) values of \( \b {_{2}} \):
\( \beta _{2}=2.35 \) (the minimum value allowed at one sigma by
SNIa \cite{Dalal}), and \( \b {_{2}}=4.02 \) (the best fit to the
SNIa data \cite{Dalal})). These two reference values will be used
in all the following discussion.}{\small \par}

{\small We shall first illustrate the \( \sg _{8} \) constraint in
Fig. 1 by plotting the curves corresponding to the experimental values
(\ref{exp}) (with a \( 3 \) sigma error band) in the plane \( \{z_{f},\, \beta _{1}\} \)
with \( \beta _{2} \) fixed . The upper curves (and the darker regions)
corresponds to lower values of \( \sigma _{8}. \) In this figure
(and in the next one) the white region has been excluded because the
transfer function parameter \( \Omega _{c(\Lambda cdm)} \) becomes
larger than unity. The dashed horizontal lines represent the upper
bound on \( z_{f} \) imposed by the baryon constraint (\ref{barfreez}).
The allowed region is below the dashed line, and within the upper
and lower white curves. In the left panel of Fig. 1 we have used \( \beta _{2}=2.35 \),
and we obtain that the maximum past-extension of the accelerated regime
is \begin{equation}
\label{zfzb}
z_{f}\leq z_{b}\simeq 5.07.
\end{equation}
 In the right panel, for \( \b {_{2}}=4.02 \), we obtain \( z_{f}\leq z_{b}\simeq 3.47. \)
At the two-sigma lower limit, \( \beta _{2}=1.55 \), the acceleration
extends to \( z_{f}\simeq 8. \)}{\small \par}

{\small By contrast, it is easy to see that in dark-energy models
of more conventional type, i.e. uncoupled to dark matter, with frozen
equation of state \( w_{\phi } \) and matter density \( \Omega _{m} \),
the acceleration starts at the redshift \begin{equation}
\label{zacc}
z_{acc}=[(3w_{\phi }-2)(\Omega _{m}-1)/\Omega _{m}]^{1/(3-3w_{\phi })}-1.
\end{equation}
 This value, for all \( w_{\phi }<2/3 \) (i.e. for a present accelerated
regime), is always smaller than unity if \( \Omega _{m}=0.3\pm 0.1 \).
Therefore, a (future) unambiguous measurement of the expansion rate
at \( z>1 \) could be a powerful method to distinguish between coupled
and uncoupled models of dark energy.}
\begin{figure}
{\small \vskip 3mm }{\small \par}

{\centering \resizebox*{!}{8cm}{\includegraphics{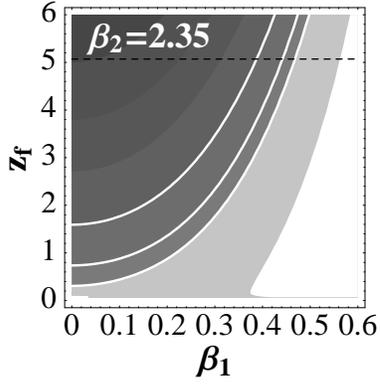}} \par}

{\centering \resizebox*{!}{8cm}{\includegraphics{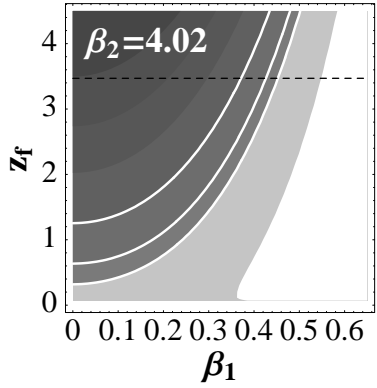}} \par}

\caption{\textsl{\small Curves at constant \protect\protect\( \sg _{8}\protect \protect \)
(from eq. (\ref{relazione})), and \protect\protect\( \b {_{2}}\protect \protect \)
fixed at the values 2.35 (top) and \protect\protect\( 4.02\protect \protect \)
(bottom). The allowed region is below the dashed line, and within the
upper and lower white curves, corresponding to the limiting values
of eq. (\ref{exp}). }}

{\small \label{f1}}{\small \par}
\end{figure}
{\small \par}

{\small The plots of Fig. 1 refer to our {}``phenomenological{}''
set of parameters, and in particular to the duration of the freezing
epoch (possibly relevant to the solution of the coincidence problem).
The \( \sg _{8} \) constraint provides however interesting information
also on the set of {}``fundamental{}'' parameters of the string
effective action (\ref{21}), used for our model of dilatonic dark
energy. }{\small \par}

{\small By eliminating \( c_{1}/c_{2}=\sqrt{2/3}\mu _{2} \), and
fixing \( q_{0} \), i.e. \( \b {_{2}} \), as before, we can plot
indeed the \( \sg _{8} \) constraint in the plane spanned by the
variables \( c^{2} \) and \( V_{0} \). The result is shown in Fig.
2, again for \( \b {_{2}}=2.35 \) (left panel) and for \( \b {_{2}}=4.02 \)
(right panel). We have restored the required Planck length factors,
and given the potential \( V_{0} \) in units of \( {\textrm{eV}}^{4} \).
The allowed region is within the upper and lower white curves, as
before. The lower bound on \( c^{2} \) derived from \( \b {_{1}} \),
at fixed \( \b {_{2}} \), is satisfied for all the range of values
illustrated in the picture. }
\begin{figure}
{\small \vskip 3mm }{\small \par}

{\centering \resizebox*{!}{10cm}{\includegraphics{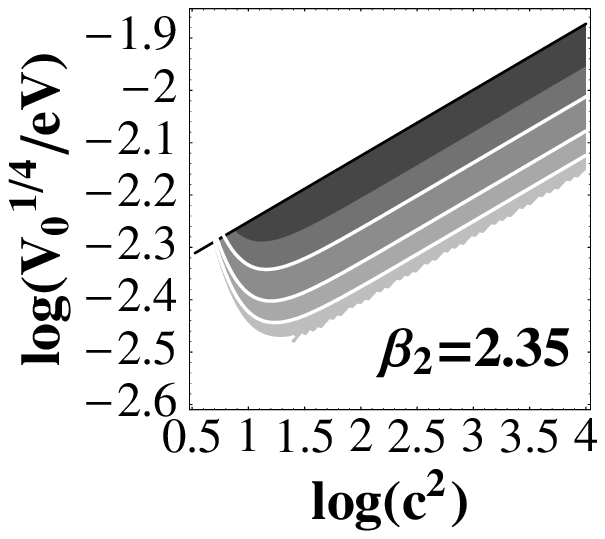}} \par}

{\centering \resizebox*{!}{10cm}{\includegraphics{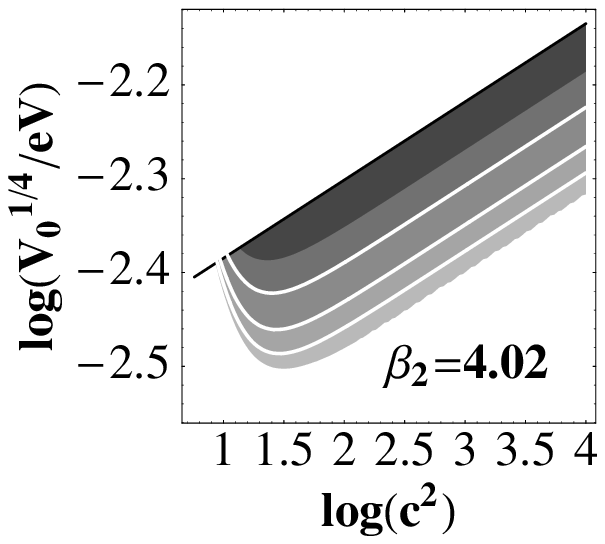}} \par}

\caption{\textsl{\small Curves at constant \protect\( \sigma _{8}\protect \),
and \protect\protect\( \b {_{2}}\protect \protect \) fixed at the
values 2.35 (top) and \protect\protect\( 4.02\protect \protect \)
(bottom). The allowed region is below the straight line representing
the consistency bound \protect\protect\( z_{f}<z_{b}\protect \protect \),
and within the upper and lower white curves corresponding to the limiting
values of eq. (\ref{exp}).}}

{\small \label{f2}}{\small \par}
\end{figure}
{\small \par}

{\small We note that the values of parameters used in \cite{GPV01}
for a particular numerical integration of the string cosmology equations
are well compatible with the above bounds. It is also important to
stress that, since the value of \( c^{2} \) is naturally expected
in the range \( 10^{2} \)--\( 10^{3} \) \cite{GPV01}, the allowed
mass scale of the dilaton potential, \( V_{0}^{1/4} \), turns out
to be fixed in a rather narrow region around (\( 10^{-2} \)--\( 10^{-3} \))
eV, even in the case of an early start of the freezing epoch. This
result confirms that, as already pointed out in \cite{GPV01}, a realistic
dark-energy scenario seems to require a rather high degree of accuracy
in determining the scale of the dilaton potential. This (fine-tuning?)
aspect of the potential is however a common problem of all scalar-field
models of dark energy. }{\small \par}

{\small It should be mentioned, to conclude this section, that we
have performed an additional observational test of our class of dilatonic
dark-energy models by comparing with the COBE data the slope of the
low-multipole \( C_{\ell } \) spectrum induced by the SW and ISW
effect. Using a simple Gaussian likelihood distribution we have concluded
that, at the confidence level of two sigma, the predicted slopes are
compatible with the data for all the allowed region of parameter space,
so that no (significant) additional constraints are generated. }{\small \par}

\section{{\small The farthest supernova}}

{\small \label{Sec4bis}}{\small \par}

{\small As a last check on the viability of the present dilatonic
dark-energy model we have considered the constraints imposed by the
most distant Type Ia supernova \cite{super} known so far, SN1997ff,
for which a very recent assessment of lensing magnification has increased
the apparent magnitude by \( 0.34\pm 0.12 \) mag \cite{super1}.
This leads to a final distance modulus (i.e. to a difference of apparent
and absolute magnitude) of \( m-M=45.49\pm 0.34 \) mag. }{\small \par}

{\small Using the definition adopted in  \cite{super} , this result
can be expressed in the following way. The luminosity distance is\begin{equation}
\label{distlum}
d_{L}=(1+z)\int _{0}^{z}\frac{dz'}{H(z')},
\end{equation}
from which one defines\begin{equation}
\Delta (m-M)=5[\log _{10}(d_{L}(z))-\log _{10}(d_{L\circ }(z))],
\end{equation}
where \( d_{L\circ }(z)=z(z+2)/(2H_{0}) \) is the luminosity distance
for Milne's model, i.e. an hyperbolic empty universe (\( \Omega _{m}=\Omega _{\Lambda }=0 \)).
For the supernova SN1997ff one obtains \( \Delta (m-M)\approx -0.15\pm0 .34 \)
at \( z\approx 1.755 \), in good agreement with a \( \Lambda  \)CDM
model characterized today by \( \Omega _{m}=0.35,\Omega _{\Lambda }=0.65 \)
\cite{super1}. For such a model, the Universe at \( z=1.755 \) is
already well within the decelerated epoch, which starts around \( z=0.548 \)
(see eq. \ref{zacc}). }{\small \par}

{\small This does not imply, however, that all models which are accelerated
at large \( z \) are ruled out (even without mentioning the still
unclear experimental uncertainties of such a supernova detection).
Let us calculate indeed the luminosity-distance along a stationary
regime \( \rho _{m}\sim \rho _{\phi } \), for a spatially flat geometry.
Neglecting for the moment the baryon contribution, the Friedmann equation
is \[
H^{2}=H_{0}^{2}\left[ \Omega _{m}(a/a_{0})^{-3w}+\Omega _{\phi }(a/a_{0})^{-3w}\right] =H_{0}^{2}(a/a_{0})^{-3w},\]
 where, in particular, \( w=w_{2} \) (see eq. \ref{230}) for our
freezing regime. The corresponding luminosity-distance (for \( w\not =2/3) \)
is: \begin{equation}
\label{dl}
d_{L}=(1+z)\int _{0}^{z}\frac{dz'}{H(z')}=\frac{2(1+z)}{(2-3w)H_{0}}\left[ (1+z)^{-\frac{3w}{2}+1}-1\right] .
\end{equation}
 For an accelerated evolution with \( w<2/3 \) (i.e., in our case,
\( q_{0}>1 \)), and for large \( z \), we have \( d_{L}\sim z^{2-3w/2} \),
while for Milne's cosmology \( d_{L}\sim z^{2}. \) It follows that,
at large \( z \) and for any \( w>0 \), the Milne model always provides
larger distances (and thus larger apparent magnitudes) than a model
of stationary evolution. As a consequence, a negative value of \( \Delta (m-M) \),
referred to Milne, does not necessarily corresponds to deceleration. }{\small \par}

{\small For a more precise illustration of this important point we
have plotted in Fig. 3 the distance modulus \( \Delta (m-M) \) for
the accelerated freezing phase of our dilatonic dark-energy model.
We have numerically integrated the luminosity-distance functions,
including baryons, for the two particular values \( \beta _{2}=2.35,\beta _{2}=4.02 \)
already used in the previous figures (it may be useful to recall that,
when \( \Om _{c} \) is fixed to \( 0.3 \), these two values of \( \b {_{2}} \)
correspond to \( w_{2}=0.5 \) and \( w_{2}=0.4 \), respectively).
It can be seen from the picture that, in both cases, the curves representing
the cosmic evolution of our model,} \emph{\small although deeply inside
the accelerating regime}{\small , remain well within one sigma from
the (lensing-corrected) SN1997ff data, while providing, at the same
time, a reasonable fit of the binned data of all the other supernovae. }{\small \par}

\begin{figure}
{\centering \includegraphics{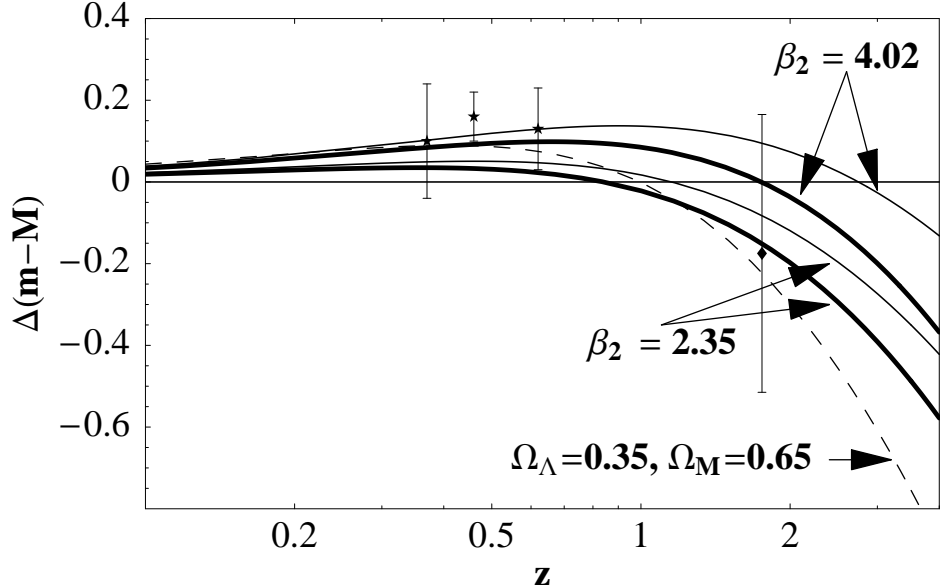} \par}

\caption{\textsl{\small The distance modulus \protect\protect\( m-M\protect \protect \),
referred to Milne's cosmology, for a \protect\protect\( \Lambda \protect \protect \)CDM
model (dashed curve) and for dilatonic dark-energy model during the
freezing epoch, for two values of \protect\protect\( \beta _{2}\protect \protect \)
(thick curves when baryons are included, thin curves without baryons).
The points at \protect\protect\( z<1\protect \protect \) represent
the binned data from all the high-redshift Ia supernovae known so
far. The datum plotted at \protect\protect\( z=1.755\protect \protect \)
represents the lensing-corrected SN1997ff . All data are derived from
\cite{super,super1}.}}
\end{figure}

\section{{\small Conclusion}}

{\small \label{Sec5}}{\small \par}

{\small In this paper we have considered a phenomenological model
of dark-energy-dark-matter interactions based on the infinite bare-coupling
limit of the superstring effective action. The dilaton, rolling down
an exponentially suppressed potential, plays the role of the cosmic
field responsible for the observed acceleration, and drives the Universe
towards a final configuration dominated by a comparable amount of
kinetic, potential and CDM energy density. }{\small \par}

{\small The effective dilatonic coupling to dark matter switches on
at late enough times (i.e., large enough bare coupling), and affects
in a significant way the post-equivalence cosmological evolution.
The time-dilution of the dark-matter density, in particular, is first
slightly enhanced (during the dragging phase) and then considerably
damped (during the freezing phase) with respect to the standard \( a^{-3} \)
decay law. The large-angle fluctuation scales relevant to the observed
CMB anisotropies reenter the horizon during the dragging epoch, and
exit the horizon again during the freezing epoch. In spite of this
unconventional evolution, the growth of the matter-density perturbations
may be large enough to match consistently present observations. }{\small \par}

{\small The predicted value of the (smoothed out) density contrast
\( \sg _{8} \), compared with data obtained from cluster abundance,
defines a significant allowed region in the parameter space of the
given class of dilatonic dark-energy models. The analysis of such
an allowed region provides two main results. }{\small \par}

{\small The first is that the bounds on the past-time extension of
the accelerated (freezing) epoch are significantly weaker than in
conventional dark-energy models (uncoupled to dark matter, with frozen
equation of state). The establishment of the freezing regime, in our
class of dilatonic models, is allowed long before the present epoch
(up to \( z\simeq 5 \)), thus providing (in principle) a further relaxation
of the coincidence problem, by extending the present cosmological
configuration not only in the far future, but also towards the past. }{\small \par}

{\small The possibility of very early (\( z>1 \)) accelerated evolution
is indeed a typical signature of such a class of dilatonic models,
useful in principle to discriminate it from other (uncoupled) dark
energy models, hopefully on the grounds of future observational data.
It is important to stress, to this respect, that the farthest type
Ia supernova so far observed is at \( z\simeq 1.7 \), and is perfectly
compatible with an accelerated Universe already at that epoch, provided
the data of the magnitude-redshift diagram are consistently fitted
by the accelerated kinematics of dilatonic models. }{\small \par}

{\small The second results concerns the parameters of the (non-perturbative)
dilaton potential appearing in the strong-bare-coupling regime of
the string effective action. The dilaton mass scale \( V_{0} \),
for an efficient and realistic dark energy scenario, appears in such
a context to be tightly anchored to a value very near to the present
Hubble curvature scale. A small deviation of \( V_{0} \) from the
required value is enough to remove the predictions of the dilatonic
model from the region of parameter space allowed by the \( \sg _{8} \)
data. }{\small \par}

{\small This means that, under the assumption that the dilatonic models
discussed in this paper provide the correct explanation of the observed
cosmic acceleration, the measurements of the density contrast \( \sg _{8} \),
besides their obvious astrophysical importance, would also acquire
an interesting high-energy significance for providing an indirect
(parameter-dependent) measurements of the dilaton mass scale. }{\small \par}

{\small We note, finally, that astrophysical observations may provide
several additional constraints on dilatonic dark-energy models. For
instance, the clustering evolution of sources at high redshifts may
constrain directly the freezing growth exponent \( m_{2} \); as already
mentioned, the baryon bias that develops during freezing is also observable,
at least in principle \cite{bias}; finally, further constraints can
be derived from a computation of the full multipole spectrum of the
CMB radiation (see e.g. \cite{aqtp} for a recent study of the CMB
constraints on coupled dark-energy models with power-law potentials).
Preliminary results seem to confirm the conclusions of this paper,
but a detailed discussion of these new constraints is postponed to
a future work. }{\small \par}

{\small \acknowledgements
It is a pleasure to thank Gabriele Veneziano for useful discussions.
CU is supported by the PPARC grant PPA/G/S/2000/00115.}{\small \par}

\end{document}